\documentclass[aps,prb,reprint,superscriptaddress,showkeys]{revtex4-1}
\usepackage{graphicx,dcolumn,bm,xcolor,microtype,multirow,amscd,amsmath,amssymb,amsfonts,physics,wrapfig,txfonts,siunitx,float}

\usepackage[export]{adjustbox}

\usepackage[version=4]{mhchem}
\usepackage{comment}

\usepackage[
colorlinks=true,
linkcolor=blue,
citecolor=blue,
urlcolor=blue]{hyperref}

\graphicspath{{figures/}}

\newcommand{\Nel}{N}

\newcommand{\dw}{\downarrow}
\newcommand{\up}{\uparrow}

\newcommand{\eps}{\epsilon}

\newcommand{\IP}{\text{IP}}
\newcommand{\EA}{\text{EA}}
\newcommand{\HOMO}{\text{HOMO}}
\newcommand{\LUMO}{\text{LUMO}}
\newcommand{\Eg}{E_\text{g}}

\newcommand{\I}{\text{i}}

\begin{document}


\title{Exploring new exchange-correlation kernels in the Bethe-Salpeter equation: a study of the asymmetric Hubbard dimer}

\newcommand{\lcpq}{Laboratoire de Chimie et Physique Quantiques, Universit\'e de Toulouse, CNRS, UPS, France}
\newcommand{\lpt}{Laboratoire de Physique Th\'eorique, Universit\'e de Toulouse, CNRS, UPS, France}
\newcommand{\etsf}{European Theoretical Spectroscopy Facility (ETSF)}

\author{Roberto Orlando}
    \affiliation{\lcpq}
    \affiliation{\lpt}
    \affiliation{\etsf}
\author{Pina Romaniello}
    \affiliation{\lpt}
    \affiliation{\etsf}
\author{Pierre-Fran\c{c}ois Loos}
    \affiliation{\lcpq}


\begin{abstract}
The Bethe-Salpeter equation (BSE) is the key equation in many-body perturbation theory based on Green's functions to access response properties. 
Within the $GW$ approximation to the exchange-correlation kernel, the BSE has been successfully applied to several finite and infinite systems. However, it also shows some failures, such as underestimated triplet excitation energies, lack of double excitations, ground-state energy instabilities in the dissociation limit, etc.
In this work, we study the performance of the BSE within the $GW$ approximation as well as the $T$-matrix approximation for the excitation energies of the exactly solvable asymmetric Hubbard dimer. 
This model allows one to study various correlation regimes by varying the on-site Coulomb interaction $U$ as well as the degree of the asymmetry of the system by varying the difference of potential $\Delta v$ between the two sites. 
We show that, overall, the $GW$ approximation gives more accurate excitation energies than $GT$ over a wide range of $U$ and $\Delta v$. 
However, the strongly-correlated (i.e., large $U$) regime still remains a challenge.
\end{abstract}

\maketitle

\section{Introduction}
\label{sec:intro}
Spectroscopy, such as photoemission, optical absorption, and electron energy loss, represents a powerful tool to characterize materials. From the theoretical point of view, a commonly used formalism to describe spectroscopic properties of infinite and finite systems from first principles, i.e., without the use of empirical parameters, is many-body perturbation theory based on Green's functions. \cite{CsanakBook,FetterBook,Martin_2016} The one-body Green's function provides, for example, the charged excitation energies of the system, which are related to the energies measured in direct and inverse photoemission spectroscopies. In particular, one can obtain the fundamental gap $\Eg = \IP - \EA$, where $\IP$ and $\EA$ are the ionization potential and the electron affinity of the system, respectively. With the two-body Green's function, instead, one can describe the neutral (or optical) excitations of the system that are measured via absorption spectroscopy.

In practice, it is only possible to get approximate one- and two-body Green's functions through the solution of Dyson-like equations in which electron correlation is approximated. In particular, for the one-body Green's function, one usually relies on the so-called $GW$ approximation \cite{Hedin_1965} to the self-energy to compute the charged excitations, \cite{Aryasetiawan_1998,Onida_2002,Reining_2017,Golze_2019,Bruneval_2021}  where $G$ is the one-body Green's function and $W=\epsilon^{-1}v_c$ (with $\epsilon^{-1}$ and $v_c$ the inverse dielectric function and the Coulomb interaction, respectively) is the dynamically screened Coulomb potential. Within this approximation, the interactions between particles are described with an infinite summation of a particular type of diagrams called bubble diagrams, which are known to be appropriate to describe systems in the high-density limit and/or the weakly correlated regime. \cite{Gell-Mann_1957,Nozieres_1958}

For the two-body Green's function, one solves the Bethe-Salpeter equation (BSE), \cite{Salpeter_1951,Strinati_1988,Blase_2018,Blase_2020} which can be derived in the particle-hole (ph) or particle-particle (pp) channels. Depending on the targeted channel, the poles of two-body Green's function represent a particular class of bound states. Here, we select the electron-hole channel to naturally obtain the neutral excitations. By exploiting its two-body nature, the BSE can be recast into an effective non-Hermitian eigenvalue problem, whose eigenvalues are the neutral excitation energies of the system and the corresponding eigenvectors are linked to the oscillator strengths which are related to the intensity of the peaks in the absorption spectrum. In its standard implementation, this linear eigenvalue problem takes as inputs the quasiparticle energies and corresponding one-body orbitals from a previous $GW$ calculation and correlates the motion of the dressed electron and hole through the statically-screened Coulomb interaction.   

This method has proven to be very effective to study optical spectra in solids \cite{Onida_2002} and, more recently, it has also made its appearance in quantum chemistry to calculate optical excitations in molecules. \cite{Rohlfing_1999a,Horst_1999,Puschnig_2002,Tiago_2003,Rocca_2010,Boulanger_2014,Jacquemin_2015a,Bruneval_2015,Jacquemin_2015b,Hirose_2015,Jacquemin_2017a,Jacquemin_2017b,Rangel_2017,Krause_2017,Gui_2018,Blase_2018,Liu_2020,Blase_2020,Holzer_2018a,Holzer_2018b,Loos_2020e,Loos_2021,McKeon_2022} While the mean-field $GW$ approximation works well for weakly/moderately correlated systems, for strongly correlated materials and molecules, failures are clearly visible. 

Recently, some of us \cite{Loos_2022} implemented the BSE within the so-called $T$-matrix  (or Bethe-Goldstone) approximation to the self-energy.  \cite{Bethe_1957,Baym_1961,Baym_1962,Danielewicz_1984a,Danielewicz_1984b,Liebsch_1981,Bickers_1989,Bickers_1991,Katsnelson_1999,Katsnelson_2002,Zhukov_2005,vonFriesen_2010,Romaniello_2012,Gukelberger_2015,Muller_2019,Friedrich_2019,Biswas_2021,Zhang_2017,Li_2021b} Within this approximation, particles are dressed with an infinite summation of ladder diagrams, which describe well the physics at low density. \cite{Danielewicz_1984a,Danielewicz_1984b,Liebsch_1981,Springer_PRL1998,Abrams_2005,Shepherd_2014} Indeed the results from Ref.~\onlinecite{Loos_2022} suggest that, in the context of the computation of molecular excitation energies, the BSE within the $T$-matrix formalism performs best in few-electron systems where the electron density remains low. However, a more complete understanding of the observed trends is still missing. 
 
In this work, we get more insights into the $GW$ and the $T$-matrix approximations by studying them on the Hubbard dimer at half-filling. This paradigm is widely used in literature to test many-body methods. \cite{Verdozzi_1995,Schindlmayr_1998b,Pollehn_1998,vonFriesen_2009,vonFriesen_2010,Romaniello_2009a,Romaniello_2012} For instance, Carrascal \textit{et al.} studied ground-state \cite{Carrascal_2015} within Kohn-Sham (KS) density-functional theory (DFT) and excited-state properties within (time-dependent) DFT \cite{Carrascal_2018} using the asymmetric Hubbard dimer. The $GW$ \cite{Romaniello_2009a,Vanzini_2018,DiSabatino_2021} and $T$-matrix approximations \cite{Baym_1961,Baym_1962,vonFriesen_2010,Romaniello_2012,Loos_2022,Zhang_2017} have also been studied using this ubiquitous model.  Here, we use the asymmetric Hubbard dimer, which can be solved exactly and, hence, offers the possibility to understand the observed trends in the charged and neutral excitation energies within the $GW$ and the $T$-matrix approximations. In particular, we will address the quality of the triplet excitation energies within the two approximations and the occurrence of complex excitation energies in the BSE.

The present manuscript is organized as follows.
In Sec.~\ref{sec:theory}, we detail how to compute charged and neutral excitation energies at the $GW$ and $T$-matrix levels.
Section \ref{sec:hubbard} introduces the two-electron asymmetric Hubbard dimer and various other theoretical and computational details.
In Sec.~\ref{sec:results}, our results on the symmetric (Subsec.~\ref{sec:sym}) and asymmetric (Subsec.~\ref{sec:asym}) Hubbard dimer are discussed, while our conclusions are drawn in Sec.~\ref{sec:conclusions}.

\section{Theory}
\label{sec:theory}

\subsection{Charged excitations}

The one-body Green's function fulfills the following Dyson equation: \cite{CsanakBook,FetterBook,Martin_2016}
\begin{equation}
    G(12)=G_\text{HF}(12)+G_\text{HF}(13) \Sigma(34)G(42)
    \label{dyson}
\end{equation}
where $G_\text{HF}$ is the Hartree-Fock (HF) Green's function chosen as a starting point and $\Sigma$ is the correlation part of the self-energy, which takes into account all the many-body effects of the system beyond Hartree and exchange. Here $1=(\boldsymbol{x}_1,t_1)=(\boldsymbol{r}_1,\sigma_1,t_1)$ describes space, spin, and time coordinates, and integration over repeated indices is implicit throughout the paper. 
The one-body Green's function has the following Lehmann representation in the frequency domain:
\begin{equation}
\label{Eqn:spectralG}
	G(\boldsymbol{x}_1,\boldsymbol{x}_2; \omega ) = \sum_\nu \frac{ \phi_\nu(\boldsymbol{x}_1) \phi^*_\nu(\boldsymbol{x}_2) }{ \omega - \eps_\nu + \I \eta \, \text{sgn}(\eps_\nu - \mu ) }
\end{equation}
where $\mu$ is the chemical potential, $\eta$ is a positive infinitesimal, $\eps_\nu = E_\nu^{\Nel+1} - E_0^{\Nel}$ for $\eps_\nu > \mu$, and $\eps_\nu = E_0^{\Nel} - E_\nu^{\Nel-1}$ for $\eps_\nu < \mu$.
Here, $E_\nu^{\Nel}$ is the total energy of the $\nu$th excited state of the $\Nel$-electron system ($\nu = 0$ being the ground state).
In the case of single-determinant many-body wave functions (such as HF or KS), the so-called Lehmann amplitudes $\phi_\nu(\boldsymbol{x})$ reduce to one-body orbitals and the poles of the Green's function $\eps_\nu$ to one-body orbital energies.

The $GW$ approximation relies on the (two-point) dynamically-screened Coulomb potential $W$ computed from a ph-random-phase approximation (ph-RPA) problem, \cite{Hedin_1965,Aryasetiawan_1998,Onida_2002,Martin_2016,Reining_2017,Golze_2019} which reads
\begin{equation}
\begin{pmatrix} 
 \boldsymbol{A}^{\text{ph}} & \boldsymbol{B}^{\text{ph}} \\
-\boldsymbol{B}^{\text{ph}} & -\boldsymbol{A}^{\text{ph}} 
\end{pmatrix}
\begin{pmatrix} 
\boldsymbol{X}^{N}_{m} \\
\boldsymbol{Y}^{N}_{m} 
\end{pmatrix} = 
\Omega_{m}^{N}
\begin{pmatrix} 
\boldsymbol{X}^{N}_{m}  \\
\boldsymbol{Y}^{N}_{m}  
\end{pmatrix}
\end{equation}
where
\begin{subequations}
\begin{align}
    A^{\text{ph}}_{ia, jb} &= (\epsilon_{a}-\epsilon_{i}) \delta_{ij}\delta_{ab} +  \bra{ib}\ket{aj} \\
    B^{\text{ph}}_{ia, jb} &=  \bra{ij}\ket{ab}
\end{align}
\end{subequations}
and
\begin{equation}
    \braket{pq}{rs} = \iint d\boldsymbol{x}_{1}d\boldsymbol{x}_{2} \chi^{*}_{p}(\boldsymbol{x}_{1})\chi^{*}_{q}(\boldsymbol{x}_{2})\frac{1}{\abs{\boldsymbol{r}_{1}-\boldsymbol{r}_{2}}}\chi_{r}(\boldsymbol{x}_{1})\chi_{s}(\boldsymbol{x}_{2})
\end{equation}
The $\chi$'s are (one-body) spin orbitals such that $\chi_{p}(\boldsymbol{x})=\psi_{p}(\boldsymbol{r}) \alpha(\sigma)$ or $\chi_{p}(\boldsymbol{x})=\psi_{p}(\boldsymbol{r}) \beta(\sigma)$ [where $\psi_{p}(\boldsymbol{r})$ is a spatial orbital]. Indices $p,q,r,s$ denote occupied or unoccupied orbitals, $i,j,k,l$ are occupied orbitals, and $a,b,c,d$ are unoccupied orbitals. Moreover, $m$ labels single excitations, while $n$ labels double electron attachments or double electron detachments.

The explicit expression of the elements of the $GW$ self-energy \cite{Hedin_1965,Aryasetiawan_1998,Onida_2002,Reining_2017,Golze_2019,Bruneval_2021}
\begin{equation}
\label{eq:SigGW}
    \Sigma^{GW}(12) = i G(12) W^\text{c}(12) 
\end{equation}
(where $W^\text{c}=W-v_c$ is the correlation part of $W$) is then given by
\begin{equation}
    \Sigma^{GW}_{pq}(\omega) = \sum_{im}  \frac{\braket*{pi}{\chi_{m}^{N}}\braket*{qi}{\chi_{m}^{N}}}{\omega-\epsilon_{i}+\Omega^{N}_{m}-i\eta} + \sum_{am}  \frac{\braket*{pa}{\chi_{m}^{N}}\braket*{qa}{\chi_{m}^{N}}}{\omega-\epsilon_{a}-\Omega^{N}_{m}+i\eta}
\end{equation}
with
\begin{equation}
    \braket*{pq}{\chi_{m}^{N}} = \sum_{ia} \braket{pi}{qa}(X_{ia,m}^{N}+Y_{ia,m}^{N})
\end{equation}
 
In the $T$-matrix approximation, one has to solve the pp-RPA equations \cite{Schuck_Book,vanAggelen_2013,Peng_2013,Scuseria_2013,Yang_2013,Yang_2013b,vanAggelen_2014,Yang_2014a,Zhang_2015,Zhang_2016,Bannwarth_2020}
\begin{equation}
\begin{pmatrix}  \boldsymbol{A}^{\text{pp}} & \boldsymbol{B}^{\text{pp}} \\
-\boldsymbol{B}^{\text{pp}} & -\boldsymbol{C}^{\text{pp}} 
\end{pmatrix}
\begin{pmatrix} 
X^{N \pm 2}_{m} \\
Y^{N \pm 2}_{m} 
\end{pmatrix} = 
\Omega^{N \pm 2}_{m}
\begin{pmatrix} 
X^{N \pm 2}_{m}  \\
Y^{N \pm 2}_{m}  
\end{pmatrix}
\end{equation}
where
\begin{subequations}    
\begin{align}    
    A^{\text{pp}}_{ab, cd} &= (\epsilon_{a}+\epsilon_{b}) \delta_{ac}\delta_{bd} + \mel{ab}{}{cd}
    \\
    B^{\text{pp}}_{ab,ij} &= \mel{ab}{}{ij}
    \\   
    C^{\text{pp}}_{ij,kl} &= -(\epsilon_{i}+\epsilon_{j}) \delta_{ik}\delta_{jl}+\mel{ij}{}{kl}
\end{align}
\end{subequations}
and $\mel{pq}{}{rs} = \braket{pq}{rs} - \braket{pq}{sr}$.

The elements of the $GT$ (or $T$-matrix) self-energy 
\begin{equation}
\label{eq:SigGT}
    \Sigma^{GT}(12) = i \int d34 G(43) T^\text{c}(13;24)
\end{equation}
(where $T^\text{c}$ is the correlation part of the $T$-matrix) are given by
\begin{equation}
    \Sigma^{GT}_{pq}(\omega) = \sum_{in} \frac{\braket*{pi}{\chi^{N+2}_{n}}\braket*{qi}{\chi^{N+2}_{n}}}{\omega+\epsilon_{i}-\Omega^{N+2}_{n}+i\eta} +
    \sum_{an}  \frac{\braket*{pa}{\chi^{N-2}_{n}}\braket*{qa}{\chi^{N-2}_{n}}}{\omega+\epsilon_{a}-\Omega^{N-2}_{n}-i\eta}
\end{equation}
where
\begin{equation}  
    \braket*{pq}{\chi_{n}^{N \pm 2}} = \sum_{c<d}\mel{pq}{}{cd}X_{cd,n}^{N \pm 2}+\sum_{k<l}\mel{pq}{}{kl}Y_{kl,n}^{N \pm 2} 
    \label{Sigma bielec aa}
\end{equation}

Next, the quasi-particle energies are obtained by linearizing the self-energy \cite{Strinati_1980,Hybertsen_1985a,Hybertsen_1986,Godby_1988,Linden_1988,Northrup_1991,Blase_1994,Rohlfing_1995,Shishkin_2007}
\begin{equation}
    \epsilon^\text{QP}_{p} = \epsilon_{p}+Z_{p} \Sigma_{pp}(\omega = \epsilon_{p})
    \label{QPp}
\end{equation}
with $\Sigma_{pp} = \Sigma^{GW}_{pp}$ or $\Sigma^{GT}_{pp}$, and where 
\begin{equation}
    Z_{p} = \qty[ 1-\eval{\pdv{\Sigma_{pp}(\omega)}{\omega}}_{\omega= \epsilon_{p}} ]^{-1}
\end{equation} 
is the renormalization factor that provides the spectral weight associated with the quasiparticle solution.

\subsection{Neutral excitations}

The Bethe-Salpeter equation reads \cite{Salpeter_1951,Strinati_1988,Blase_2018,Blase_2020}
\begin{multline}
L(12;1^{\prime}2^{\prime}) = L_{0}(12;1^{\prime}2^{\prime}) 
\\ 
+ \int d3456 L_{0}(14;1^{\prime}3) \Xi(35;46) L(62;52^{\prime})
\label{bse}
\end{multline}
with
\begin{equation}
    L_{0}(12;1^{\prime}2^{\prime}) = -iG(12^{\prime})G(21^{\prime})
    \label{L0}
\end{equation}
The kernel $\Xi(35;46) = \fdv*{\Sigma(34)}{G(65)}$ is a two-body effective interaction and $L$ is the bound part of the two-body Green's function $G_{2}$ such that
\begin{equation}
    iL(12;1'2') =-G_{2}(12;1'2') +G(11')G(22') 
    \label{L}
\end{equation}
From Eq.~\eqref{bse} and within the static approximation, \cite{Strinati_1988,Romaniello_2009b,Sangalli_2011,Loos_2020h} we obtain the following eigenvalue equation
\begin{equation}
\label{eq:BSE}
    \begin{pmatrix} 
         \boldsymbol{A}^{\text{BSE}} & \boldsymbol{B}^{\text{BSE}} \\
        -\boldsymbol{B}^{\text{BSE}} & -\boldsymbol{A}^{\text{BSE}} 
    \end{pmatrix}
    \begin{pmatrix} 
        \boldsymbol{X}^{\text{BSE}}_{m} \\
        \boldsymbol{Y}^{\text{BSE}}_{m} 
    \end{pmatrix} = 
    \Omega_{m}^{\text{BSE}}
    \begin{pmatrix} 
        \boldsymbol{X}^{\text{BSE}}_{m}  \\
        \boldsymbol{Y}^{\text{BSE}}_{m} 
    \end{pmatrix}
\end{equation}
with
\begin{subequations}
\begin{align}
    \label{eq:A_BSE}
    A^{\text{BSE}}_{ia, jb} &= (\epsilon^\text{QP}_{a}-\epsilon^\text{QP}_{i}) \delta_{ij}\delta_{ab} + \mel{ib}{}{aj} - \Xi^\text{c}_{ij,ba}(\omega=0) 
    \\
    \label{eq:B_BSE}
    B^{\text{BSE}}_{ia, jb} &=  \mel{ij}{}{ab} - \Xi^\text{c}_{ib,ja}(\omega=0)
\end{align}
\end{subequations}
and where $\Omega_{m}^{\text{BSE}}$ is the $m$th BSE neutral excitation, the $\epsilon^\text{QP}_{p}$'s (with QP = $GW$ or $GT$) are the quasiparticle energies previously obtained, and the correlation kernel is $\Xi^\text{c} = W^\text{c}$ or $-T^\text{c}$, with
\begin{multline}
\label{eq:Wc}
    W^\text{c}_{ia,jb}(\omega) 
    = \sum_{m} \braket*{ij}{\chi_m^{N}}\braket*{ab}{\chi_m^{N}} 
    \\
    \times \qty(\frac{1}{\omega-\Omega^{N}_{m}+i\eta}-\frac{1}{\omega+\Omega^{N}_{m}-i\eta})
\end{multline}
and
\begin{equation}
\label{eq:Tc}
T^\text{c}_{ia,jb}(\omega) 
    = \sum_{n} \frac{\braket*{ib}{\chi^{N+2}_{n}}\braket*{aj}{\chi^{N+2}_{n}}}{\omega-\Omega^{N+2}_{n}+i\eta} 
    - \sum_{n} \frac{\braket*{ib}{\chi^{N-2}_{n}}\braket*{aj}{\chi^{N-2}_{n}}}{\omega-\Omega^{N-2}_{n}-i\eta}
\end{equation}
Note that, due to the frequency-independent nature of the static BSE, one cannot access double (and higher) excitations.
\cite{Strinati_1988,Romaniello_2009b,Rohlfing_2000,Ma_2009a,Ma_2009b,Loos_2019,Sangalli_2011,Loos_2020h,Authier_2020,Monino_2021}
In the following, we set $\eta = 0$.

\section{Hubbard dimer}
\label{sec:hubbard}

The Hamiltonian of the asymmetric Hubbard dimer is
\begin{equation}
    \hat{H} = 
    - t \sum_{\sigma=\up,\dw} \qty( \hat{a}_{1\sigma}^{\dagger}\hat{a}_{2\sigma}+\hat{a}_{2\sigma}^{\dagger}\hat{a}_{1\sigma} )
    + U \sum_{i=1}^{2} \hat{n}_{i\up} \hat{n}_{i\dw} 
   +  \Delta v \frac{\hat{n}_{2} - \hat{n}_{1}}{2}
    \label{H dim as}
\end{equation}
where $t > 0$ is the hopping parameter, $U \ge 0$ is the local Coulomb interaction, $\hat{n}_{i\sigma} = \hat{a}_{i\sigma}^{\dagger}\hat{a}_{i\sigma}$ is the spin density operator on site $i$, $\hat{n}_{i} = \hat{n}_{i\uparrow} + \hat{n}_{i\downarrow}$ is the density operator on site $i$, and $\Delta v = v_{1} - v_{2}$ (with $v_{1} > v_{2}$ and $v_1 + v_2 = 0$) is the potential difference between the two sites. 
The operator $\hat{a}_{i\sigma}^{\dagger}$ ($\hat{a}_{i\sigma}$) creates (annihilates) an electron of spin $\sigma$ on site $i$.
In the following, all quantities are reported in reduced units or, equivalently, in units of $t$.

For $N=2$, expanding the Hamiltonian in the site basis $\ket{\up_{1} \dw_{2}}$, $\ket{\dw_{1} \up_{2}}$, $\ket{0_{1}\up\dw_{2}}$, and $\ket{\up\dw_{1}0_{2}}$, the triplet excited-state wave function (with $M_S = 0$) is
\begin{equation}
    \ket*{{}^3\Psi^{N}} = \frac{\ket{\up_{1} \dw_{2}} + \ket{\dw_{1} \up_{2}}}{\sqrt{2}},  
    \label{Psi dimer as} 
\end{equation}  
while the wave functions of the singlet ground ($n = 0$) and excited ($n = 1$ and $2$) states have the form
\begin{equation}
    \ket*{{}^1\Psi_{n}^{N}} 
    = c_{1n}\ket{\up_{1} \dw_{2}}
    + c_{2n}\ket{\dw_{1} \up_{2}} 
    + c_{3n}\ket{0_{1}\up\dw_{2}}  
    + c_{4n}\ket{\up\dw_{1}0_{2}} 
\end{equation}  
with 
\begin{subequations}
\begin{align}
    c_{1n} & = -c_{2n} = \frac{1}{\mathcal{N}_{n}}
    \\
    c_{3n} & = \frac{1}{\mathcal{N}_{n}} \frac{2t}{U-\Delta v - E_{n}^{N}}
    \\
    c_{4n} & = \frac{1}{\mathcal{N}_{n}} \frac{2t}{U+\Delta v - E_{n}^{N}}   
\end{align}
\end{subequations}
and
\begin{equation}
    \mathcal{N}_{n} = \sqrt{2+\qty(\frac{2t}{U-\Delta v - E_{n}^{N}})^{2}+\qty(\frac{2t}{U+\Delta v - E_{n}^{N}})^{2}}
\end{equation}
where the $E_{n}^{N}$'s are the corresponding (exact) eigenenergies of Eq.~\eqref{H dim as}.
The differences between these $N$-electron energies allow us to compute the exact neutral excitation energies of the system, while the exact IP and EA are computed as energy differences with respect to the system with $N-1$ and $N+1$ electrons. 
We note that $\ket*{{}^1\Psi_{2}^{N}} $ is the excited-state wave function associated with the so-called double excitation, which we will not address in the following.

We saw in Sec.~\ref{sec:theory} that $W$ is a two-point quantity [see Eq.~\eqref{eq:SigGW}] while $T$ is a four-point quantity [see Eq.~\eqref{eq:SigGT}]. 
Because the Coulomb interaction is local in the Hubbard model, i.e., $v(1,2) = U \delta(1,2)$, $T$ also becomes a two-point quantity in this special case.

At the mean-field level, the system is composed by two orbitals: a bonding orbital $\epsilon_{\HOMO} = - \IP$ and an anti-bonding orbital $\epsilon_{\LUMO} = - \EA$. 
The approximated fundamental gap is thus $E_\text{g} = \IP - \EA = \epsilon_{\LUMO} - \epsilon_{\HOMO}$ and can be computed at the HF, $GW$, or $T$-matrix level depending on the choice of (quasi)particle energies.

Furthermore, we note ${}^1\boldsymbol{H}_\text{exc}$ and ${}^3\boldsymbol{H}_\text{exc}$ the singlet and triplet excitonic Hamiltonians of the dimer that are obtained by spin-resolving Eq.~\eqref{eq:BSE}. \cite{Angyan_2011,Bruneval_2016a,Monino_2021}
$^1\Omega$ and $^3\Omega$ are their corresponding positive eigenvalues that represent the excitation energies associated with the singlet-singlet and singlet-triplet transition, respectively. 
We also define the binding energy of the singlet and triplet excitations as $^1E_{\text{b}} = E_\text{g} - {}^1\Omega$ and $^3E_{\text{b}} = E_\text{g} - {}^3\Omega$, respectively, where $E_\text{g}$, ${}^1\Omega$, and ${}^3\Omega$ are obtained at the same level of theory (i.e., exact, $GW$ or $T$-matrix level).

From a general point of view, the excitonic Hamiltonian can be written as
\begin{equation}
    \boldsymbol{H}_\text{exc} = 
    \begin{pmatrix}
        E_\text{g} + a & b
        \\
        -b & -E_\text{g} - a
        \\
    \end{pmatrix}
\end{equation}
while the excitation energies are
\begin{equation}
    \Omega = \pm \sqrt{\qty(E_\text{g}+a)^2- b^2}
    \label{omega}
\end{equation}
but we keep only the resonant transition energies that correspond to the positive eigenvalues.

\section{Results}
\label{sec:results}

\subsection{Symmetric Hubbard dimer}
\label{sec:sym}

Let us first study the symmetric Hubbard dimer (i.e., $\Delta v = 0$) since it can be solved analytically in the case of a non-interacting or HF starting point, $G_{0}$ and $G_{\text{HF}}$, respectively, which further highlights the influence of the starting Green's function.
Before commenting on the performance of $GW$ and $GT$, let us first report some exact expressions for the symmetric Hubbard dimer at half-filling.

The exact IP, EA, and fundamental gap are
\begin{subequations}
\begin{align}
    \text{IP} & = - t - \frac{1}{2} \qty[ U - \sqrt{(4t)^2 + U^2} ]
    \\
    \text{EA} & = t - U + \frac{1}{2} \qty[ U - \sqrt{(4t)^2 + U^2} ]
    \\
    E_\text{g} & = - 2t + \sqrt{(4t)^2 + U^2}
\end{align}
\end{subequations}
It is instructive to study the small-$U$ limit of these quantities.
In particular, the fundamental gap behaves as 
\begin{equation}
    \label{eq:Eg}
    E_\text{g} = 2 t + \frac{U^2}{8 t} - \frac{U^4}{512 t^3} + \order{U^6}
\end{equation}
The exact singlet and triplet excitation energies are
\begin{subequations}
\begin{align}
    {}^1\Omega & = U - \frac{1}{2} \qty[ U - \sqrt{(4t)^2 + U^2} ]
    \\
    {}^3\Omega & = - \frac{1}{2} \qty[ U - \sqrt{(4t)^2 + U^2} ]
\end{align}
\end{subequations}
which yields a singlet-triplet gap equals to $U$.
In the small-$U$ limit, these excitation energies behave as
\begin{subequations}
\begin{align}
    \label{eq:Omega1}
    {}^1\Omega & = 2 t + \frac{U}{2} + \frac{U^2}{16 t} - \frac{U^4}{1024 t^3} + \order{U^6}
    \\
    \label{eq:Omega3}
    {}^3\Omega & = 2 t - \frac{U}{2} + \frac{U^2}{16 t} - \frac{U^4}{1024 t^3} + \order{U^6}
\end{align}
\end{subequations}

\subsubsection{Charged excitations}

\begin{figure}
    \includegraphics[width=\linewidth]{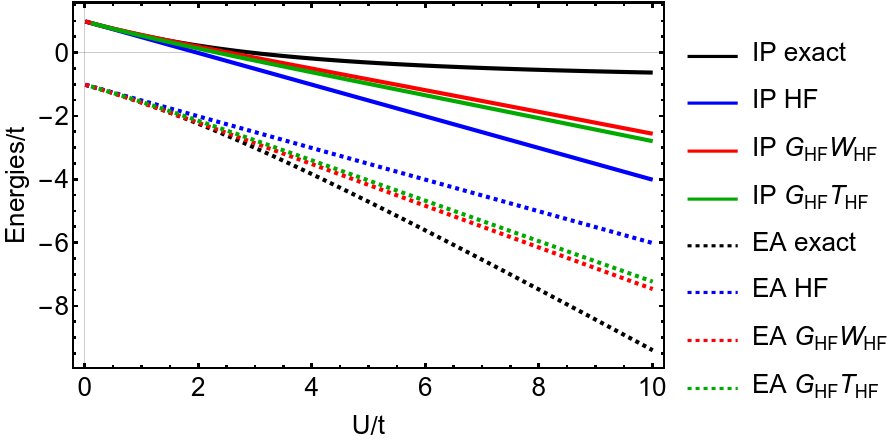}
    \caption{IPs (solid) and EAs (dashed) as functions of $U/t$ in the symmetric Hubbard dimer ($\Delta v = 0$) at various levels of theory: exact (black), HF (blue), $G_{\text{HF}}W_{\text{HF}}$ (red), and $G_{\text{HF}}T_{\text{HF}}$ (green).}
    \label{fig:QP U}
\end{figure}

We start by analyzing the charged excitation energies as functions of $U/t$. 
We are mainly interested in the quality of the quasiparticle energies (or, equivalently, the IP and EA) since they enter the BSE [see Eq.~\eqref{eq:A_BSE}]. 
In Fig.~\ref{fig:QP U}, we compare the exact IP and EA with approximate IPs and EAs computed at the $GW$ and $GT$ levels with an HF starting Green's function. 
These two schemes are coined $G_{\text{HF}}W_{\text{HF}}$ and $G_{\text{HF}}T_{\text{HF}}$ in the following.
Note that, in the specific case of the symmetric Hubbard dimer, because we linearize the self-energy [see Eq.~\eqref{QPp}], we get the same results using $G_0$ as starting point. 

In the weak correlation (i.e., small $U/t$) regime, all approximations are in good agreement with the exact results. 
In the strong correlation (i.e., large $U/t$) regime, both the $GW$ and the $T$-matrix approximations underestimate the fundamental gap quite significantly.
One would note, however, that $GW$ is slightly better than the $T$-matrix approximation. 
For comparison, in Fig.~\ref{fig:QP U}, we report also the HF results, which are clearly worse than both $G_{\text{HF}}W_{\text{HF}}$ and $G_{\text{HF}}T_{\text{HF}}$. 
These observed trends are in line with previously reported results on the same model in which the quasiparticle energies are computed without linearizing the self-energy, \cite{Romaniello_2009a,Romaniello_2012} as well as in more realistic molecular systems. \cite{Zhang_2017,Loos_2022}

Based on these results, we can study analytically the small-$U$ limits of the quantities of interest.
For example, the $G_{\text{HF}}W_{\text{HF}}$ and $G_{\text{HF}}T_{\text{HF}}$ behave as 
\begin{subequations}
\begin{align}
    E_\text{g}^{G_{\text{HF}}W_{\text{HF}}} 
    & = 2 t + \frac{U^2}{4 t}  -\frac{3 U^3}{16 t^2} + \frac{19 U^4}{128 t^3} + \order{U^5}
    \\
    E_\text{g}^{G_{\text{HF}}T_{\text{HF}}} 
    & = 2 t + \frac{U^2}{8 t}  -\frac{3 U^3}{64 t^2} + \frac{9 U^4}{512 t^3} + \order{U^5}
\end{align}
\end{subequations}
Compared with the exact gap reported in Eq.~\eqref{eq:Eg}, one can see that $G_{\text{HF}}W_{\text{HF}}$ is already wrong at second order in $U$, while the $G_{\text{HF}}T_{\text{HF}}$ gap is correct up to the quadratic term thanks to the inclusion of the second-order direct and exchange diagrams, as already mentioned in Ref.~\onlinecite{Romaniello_2012}.
However, it exhibits a spurious cubic term and the wrong quartic coefficient. 
(Note that the renormalization factor only affects the value of the quartic coefficient.)

\subsubsection{Neutral excitations}

\begin{figure}
    \includegraphics[width=\linewidth]{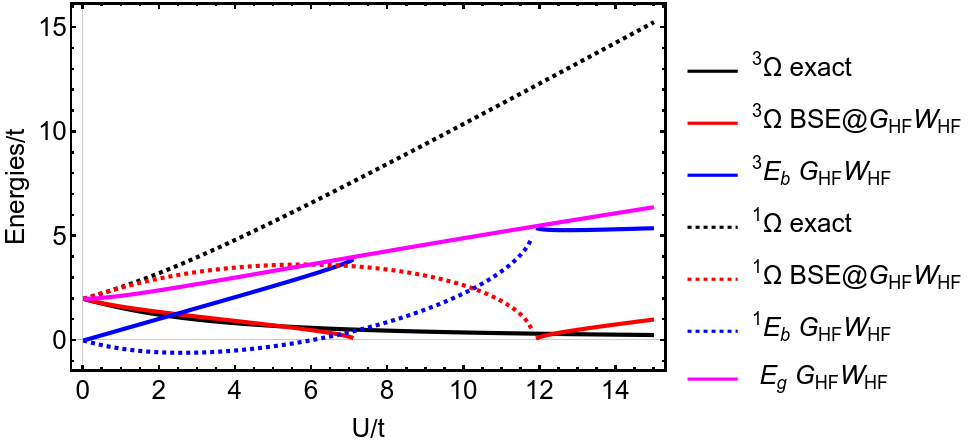}
    \caption{Neutral excitation energies (red), binding energies (blue), and fundamental gap (magenta) as functions of $U/t$ in the symmetric Hubbard dimer ($\Delta v = 0$) for the triplet (solid) and singlet (dashed)  excited states at various levels of theory: exact (black) and BSE@$G_{\text{HF}}W_{\text{HF}}$.}
    \label{fig:gwall}
\end{figure}

We now focus on the neutral excited states of the symmetric Hubbard dimer.
Within the $G_\text{HF}W_\text{HF}$ approximation to the self-energy, the excitonic Hamiltonians read
\begin{subequations}
\begin{align}
    \label{eq:H3_GW}
    {}^3\boldsymbol{H}_\text{exc}^{G_\text{HF}W_\text{HF}} & = 
    \begin{pmatrix}
        E_\text{g}^{G_\text{HF}W_\text{HF}}-\frac{U}{2}  & \frac{U}{2} \qty( \frac{U}{t+U} - 1 )
        \\
        -\frac{U}{2} \qty( \frac{U}{t+U} - 1 ) & -\qty(E_\text{g}^{G_\text{HF}W_\text{HF}}-\frac{U}{2})
        \\
    \end{pmatrix}
    \\
    \label{eq:H1_GW}
    {}^1\boldsymbol{H}_\text{exc}^{G_\text{HF}W_\text{HF}} & = 
    \begin{pmatrix}
        E_\text{g}^{G_\text{HF}W_\text{HF}}+\frac{U}{2}  & \frac{U}{2} \qty( \frac{U}{t+U} + 1 )    
        \\
        -\frac{U}{2} \qty( \frac{U}{t+U} + 1 ) & -\qty(E_\text{g}^{G_\text{HF}W_\text{HF}}+\frac{U}{2})    
        \\
\end{pmatrix}
\end{align}
\end{subequations}
When we use $G_{0}$ instead of $G_\text{HF}$ as a starting Green's function, we obtain the same excitonic Hamiltonians. 
This is due to the fact that in the electron-hole polariability, which enters the expression of $W$, the additional terms included in the HF expressions cancel. As we shall see, this is not the case for the particle-particle polarizability needed to calculate the $T$-matrix.
We see that the particle-hole polarisability induces an asymmetry in the resonant and coupling terms due to the spin structure of $W$ which implies that there is no Hartree term for the triplet state. 
The excitation energies are shown in Fig.~\ref{fig:gwall} alongside the binding energies of the singlet and triplet excitation energies.

The $GW$ approximation (red curves) describes better the exact triplet excitation energy, ${}^3\Omega$, than the exact singlet excitation energy, ${}^1\Omega$. 
Between $U/t \approx 7$ and $U/t\approx 12$, however, ${}^3\Omega^{GW}$ (solid red curve) turns complex due to a triplet instability in the BSE matrix.
This occurs because the coupling term $b=\frac{U}{2} \qty( \frac{U}{t+U} - 1 )$ becomes larger than the resonant term $E_\text{g}-a=E_\text{g}^{G_\text{HF}W_\text{HF}}+U/2$, as it is clear from Eq.~\eqref{omega}. 
Interestingly, in this same range, the $GW$ binding energy ${}^3E_\text{b}$ (solid blue curve) reaches the $GW$ fundamental gap (solid magenta line). 
Moreover, at $U/t \approx 7$, ${}^1\Omega^{GW}$ (dashed red curve) becomes lower than the fundamental gap and, hence, becomes, by definition, a ``bound'' state, while, at $U/t \approx 12$, it becomes complex (singlet instability) when ${}^1E_\text{b}$ (dashed blue curve) reaches the $GW$ fundamental gap (dashed magenta line).

Within the $G_\text{HF}T_\text{HF}$ approximation to the self-energy, the excitonic Hamiltonians read
\begin{subequations}
\begin{align}
    \label{eq:H3_GT}
    {}^3\boldsymbol{H}_\text{exc}^{G_\text{HF}T_\text{HF}} & = 
    \begin{pmatrix}
        E_\text{g}^{G_\text{HF}T_\text{HF}}-\frac{U}{2}  & \frac{U}{2} \qty( \frac{Ut/2}{h^{2}-U^{2}} - 1 )   \\
        -\frac{U}{2} \qty( \frac{Ut/2}{h^{2}-U^{2}} - 1 ) & -\qty(E_\text{g}^{G_\text{HF}T_\text{HF}}-\frac{U}{2})\\
    \end{pmatrix}
    \\
    \label{eq:H1_GT}
    {}^1\boldsymbol{H}_\text{exc}^{{G_\text{HF}T_\text{HF}}} & = 
    \begin{pmatrix}
        E_\text{g}^{G_\text{HF}T_\text{HF}}+\frac{U}{2}  & -\frac{U}{2} \qty( \frac{Ut/2}{h^{2}-U^{2}} - 1 )    \\
        \frac{U}{2} \qty( \frac{Ut/2}{h^{2}-U^{2}} - 1 ) & -\qty(E_\text{g}^{G_\text{HF}T_\text{HF}}+\frac{U}{2})\\
    \end{pmatrix}
\end{align}
\end{subequations}
with $h=\sqrt{4t^2+2Ut}$ and where one would notice a similar form as Eqs.~\eqref{eq:H3_GW} and \eqref{eq:H1_GW} for the resonant and antiresonant blocks, and a very different expression for the coupling blocks due again to the difference in spin structure between $W$ and $T$.
In the case of $G_0T_0$, one gets similar expressions except that one must replace the coupling block by $\pm \frac{U}{2} \qty( \frac{Ut/2}{h^{2}} - 1 )$ instead of $\pm \frac{U}{2} \qty( \frac{Ut/2}{h^{2}-U^{2}} - 1 )$. 

Taylor expanding the singlet and triplet $G_\text{HF}T_\text{HF}$ excitations at small $U$ shows that
\begin{subequations}
\begin{align}
    {}^1\Omega^{G_\text{HF}T_\text{HF}} & = 2 t + \frac{U}{2} + \frac{U^2}{16 t} - \frac{U^3}{64 t^2} + \frac{3 U^4}{512 t^3} + \order{U^5}
    \\
    {}^3\Omega^{G_\text{HF}T_\text{HF}} & = 2 t -\frac{U}{2} + \frac{U^2}{16 t} - \frac{3 U^3}{64 t^2} + \frac{7 U^4}{512 t^3} + \order{U^5}
\end{align}
\end{subequations}
which is correct up to second order in $U/t$ when compared to Eqs.~\eqref{eq:Omega1} and \eqref{eq:Omega3}.
Interestingly, replacing the $T$-matrix quasiparticle with the HF one-particle energies change the sign of the second-order term.
At the $G_\text{HF}W_\text{HF}$ level, we have
\begin{subequations}
\begin{align}
    {}^1\Omega^{G_\text{HF}W_\text{HF}} & = 2 t + \frac{U}{2} + \frac{3 U^2}{16 t} - \frac{19 U^3}{64 t^2} + \frac{251 U^4}{1024 t^3} + \order{U^5}
    \\
    {}^3\Omega^{G_\text{HF}W_\text{HF}} & = 2 t -\frac{U}{2} + \frac{3 U^2}{16 t} - \frac{5 U^3}{64 t^2} - \frac{5 U^4}{1024 t^3} + \order{U^5}
\end{align}
\end{subequations}
which clearly do not have the right quadratic behavior.

\begin{figure}
    \includegraphics[width=\linewidth]{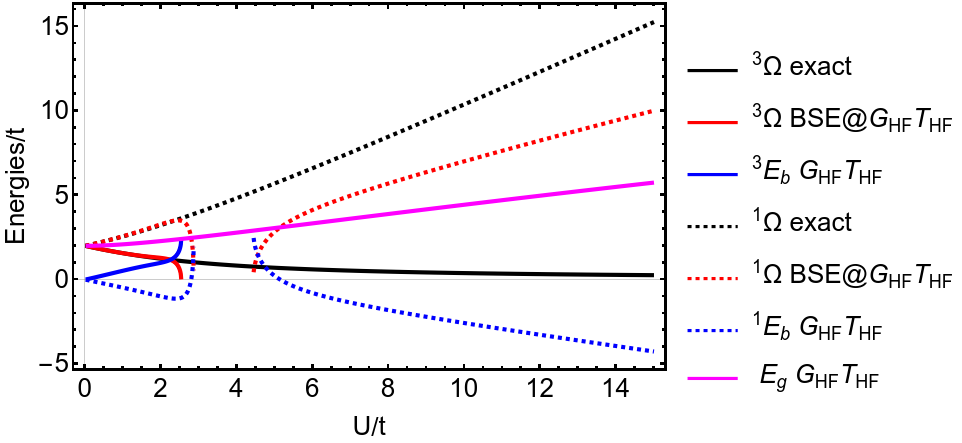}
    \includegraphics[width=\linewidth]{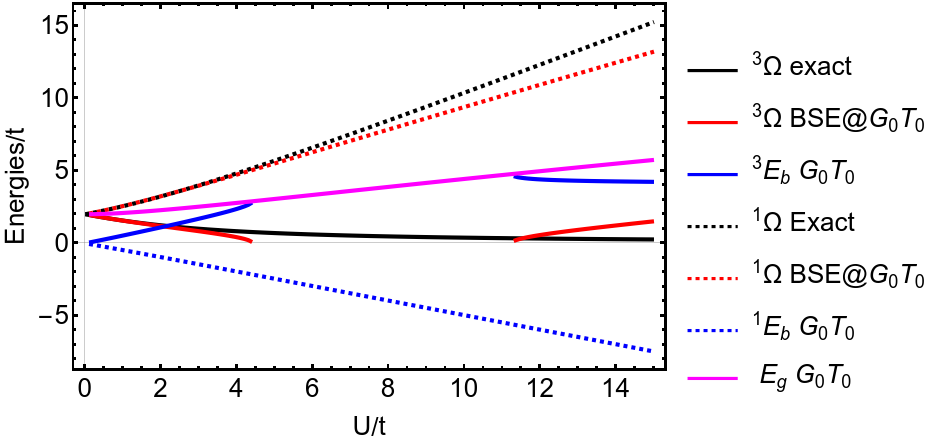}
    \caption{Neutral excitation energies (red), binding energies (blue), and fundamental gap (magenta) as functions of $U/t$ in the symmetric Hubbard dimer ($\Delta v = 0$) for the triplet (solid) and singlet (dashed) excited states at various levels of theory: exact (black), BSE@$G_{\text{HF}}T_{\text{HF}}$ (top) and BSE@$G_{0}T_{0}$  (bottom).}
    \label{fig:Oppall}
\end{figure}

The excitation energies are represented in Fig.~\ref{fig:Oppall} as functions of $U/t$.
For the case of BSE@$G_\text{HF}T_\text{HF}$ (bottom panel), the exact singlet and triplet transitions (black curves) are well described by ${}^1\Omega^{G_\text{HF}T_\text{HF}}$ and ${}^3\Omega^{G_\text{HF}T_\text{HF}}$ (red curves) until $U/t \approx 3$.
After this, they both become complex, ${}^1\Omega^{G_\text{HF}T_\text{HF}}$ becoming real again at $U/t \approx 4.5$.
As before ${}^1\Omega^{G_\text{HF}T_\text{HF}}$ and ${}^3\Omega^{G_\text{HF}T_\text{HF}}$ turn complex when their respective binding energy, ${}^1E_\text{b}$ and ${}^3E_\text{b}$ (magenta curves), reaches the value of the fundamental gap.

At the BSE@$G_0T_0$ (top panel), excitation energies are surprisingly more accurate. For example, ${}^1\Omega^{G_0T_0}$ (red dashed curve) is an excellent approximation of the exact singlet excitation energy until $U/t \approx 5$ at which it starts to deviate but remains decent.
Likewise, ${}^3\Omega^{G_0T_0}$ (solid red curve) is accurate until $U/t \approx 4$ at which a triplet instability appears.
Since the quality of the $GW$ and $GT$ quasiparticle energies as functions of $U/t$ is very similar, and since the HF contribution to the $GW$ and $GT$-based BSE kernels is the same, the difference between $GW$ and $GT$ excitation energies can be traced back to the difference in the correlation part of the $GW$- and $GT$-based kernels [see Eqs.~\eqref{eq:Wc} and \eqref{eq:Tc}, respectively].

These results show that, in order to get accurate neutral excitation energies, there is a subtle balance to fulfill between the quality of the fundamental gap and the correlation kernel.
Altering one of them will result in instabilities in the BSE problem, hence complex excitation energies.

\subsection{Asymmetric Hubbard dimer}
\label{sec:asym}
We now turn our attention to the more general case of the asymmetric dimer for which we only study $G_\text{HF}$ as a starting point.

\subsubsection{Charged excitations}
\begin{figure*}
    \centering
    \includegraphics[width=\linewidth]{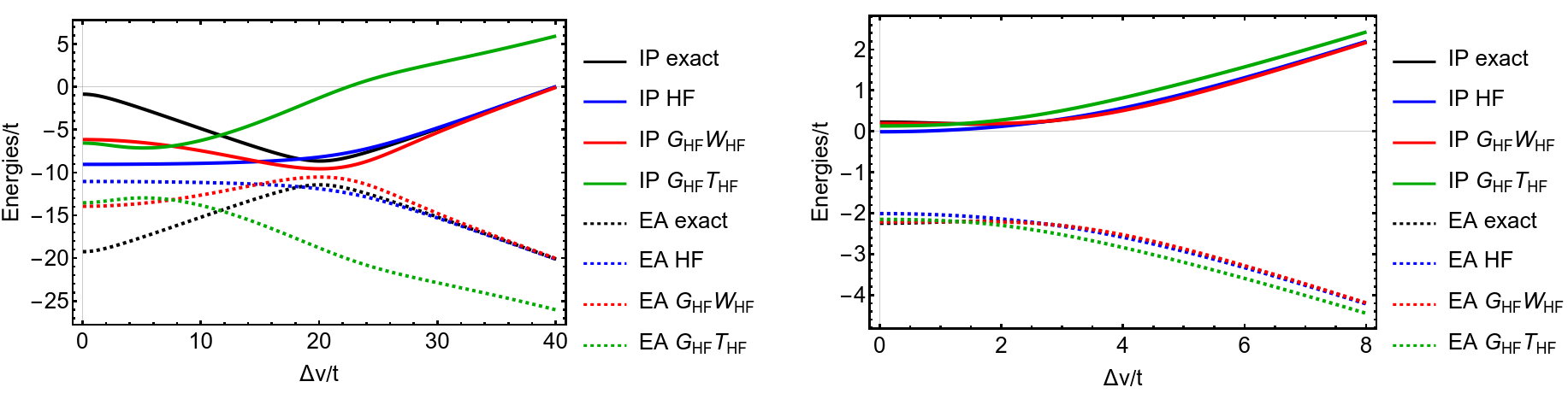}
    \caption{IP and EA as functions of $\Delta v/t$ in the asymmetric Hubbard dimer for $U/t = 20$ (left panel) and $U/t = 2$ (right panel) obtained at various levels of theory: exact (black), HF (blue), $G_\text{HF}W_\text{HF}$ (red), and $G_\text{HF}T_\text{HF}$ (green).}
    \label{fig:QP Delta v}
\end{figure*}

In Fig.~\ref{fig:QP Delta v}, we report the IP and EA as functions of $\Delta v/t$. The ratio $U/t$ is fixed at 20 so that, for $\Delta v/t<20$, we can assume to be in a strongly-correlated regime, while, for $\Delta v/t>20$, the electron correlation is weak. 
As for the symmetric case (see Sec.~\ref{sec:sym}), we show that, in the strongly-correlated regime, all approximations underestimate the exact fundamental gap, with a slightly better performance of $GW$ (red curves) as compared to $GT$ (green curves). 
For $\Delta v/t>20$, both $GW$ and HF tend to the exact values, whereas $GT$ largely deviates from the exact result. 

\begin{figure*}
    \centering
    \includegraphics[width=\linewidth]{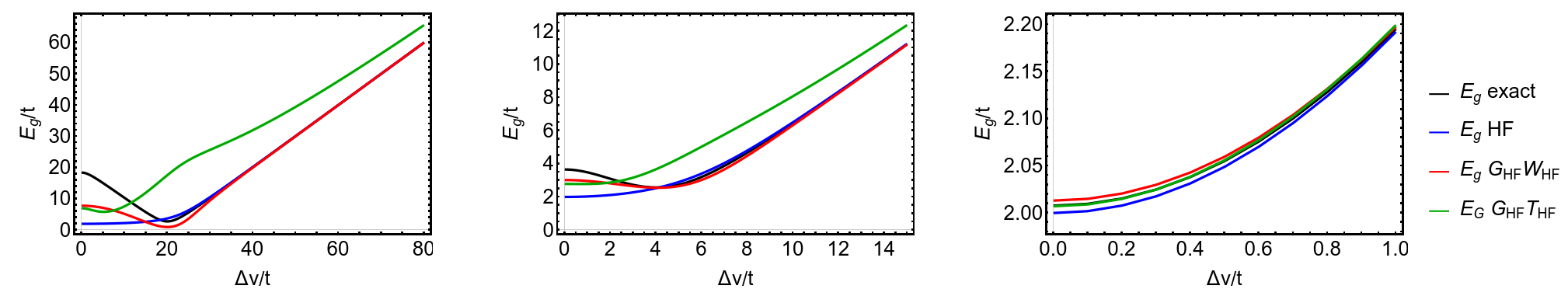}
    \caption{Fundamental gap as a function of $\Delta v/t$ in the asymmetric Hubbard dimer for $U/t = 20$ (left), $U/t = 4$ (center) and $U/t = 1/4$ (right) at various levels of theory: exact (black), HF (blue), $G_\text{HF}W_\text{HF}$ (red), and $G_\text{HF}T_\text{HF}$ (green).}
    \label{fig:Gap Delta v}
\end{figure*}

In Fig.~\ref{fig:Gap Delta v}, we analyze this trend in more detail by looking at the evolution of the fundamental gap as a function of $\Delta v/t$ for various values of $U/t$: $U/t = 20$ (left panel), $U/t = 4$ (central panel), and $U/t = 1/4$ (right panel).
We observe that the $GT$ gap merges to the exact one only for very small $U/t$.
From Fig.~\ref{fig:QP Delta v}, it becomes clear that the electron correlation is negligible for large $\Delta v/t$, which is also in line with the findings of Carrascal \textit{et al.} \cite{Carrascal_2015} on the correlation energy. 
This explains why HF (blue curves) is exact for large $\Delta v/t$. 
Moreover, $GW$ merges to HF since the screening tends to zero because of the large energy difference between the bonding and anti-bonding orbitals. 
The correlation part of the $GT$ self-energy, instead, is not negligible, and this can be attributed to the fact that the particle-particle polarizability which enters into the $T$-matrix expression (see Sec.~\ref{sec:theory}) does not go to zero in this limit.
In a future work, it would be interesting to analyze these results using concepts similar to the density-driven and functional-driven errors employed in DFT. \cite{Vuckovic_2019}

\subsubsection{Neutral excitations}
\begin{figure*}
    \includegraphics[width=0.49\linewidth]{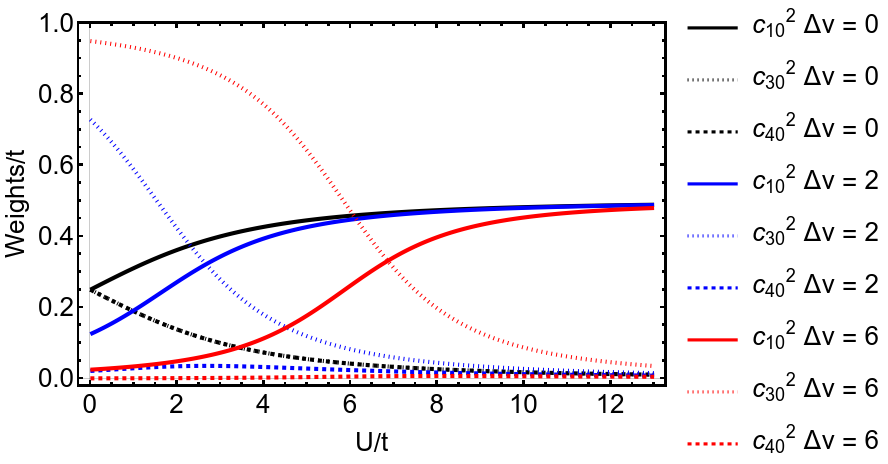}
    \includegraphics[width=0.49\linewidth]{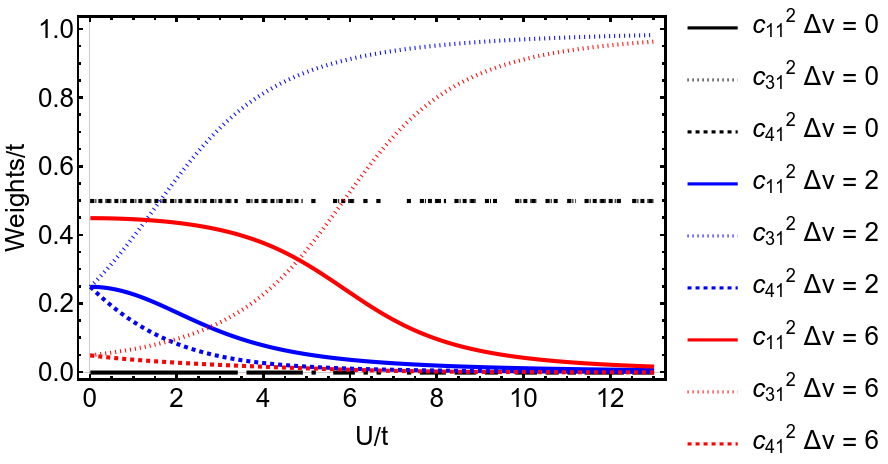}
    \caption{Weights of $^1\Psi_{0}^{N}$ (left) and $^1\Psi_{1}^{N}$ (right) as functions of $U/t$. For $t = 1$, $c_{20}^2 = c_{10}^2$ and $c_{22}^2 = c_{12}^2$.}
    \label{fig: Poids}
\end{figure*}

\begin{figure*}
    \includegraphics[scale=0.36]{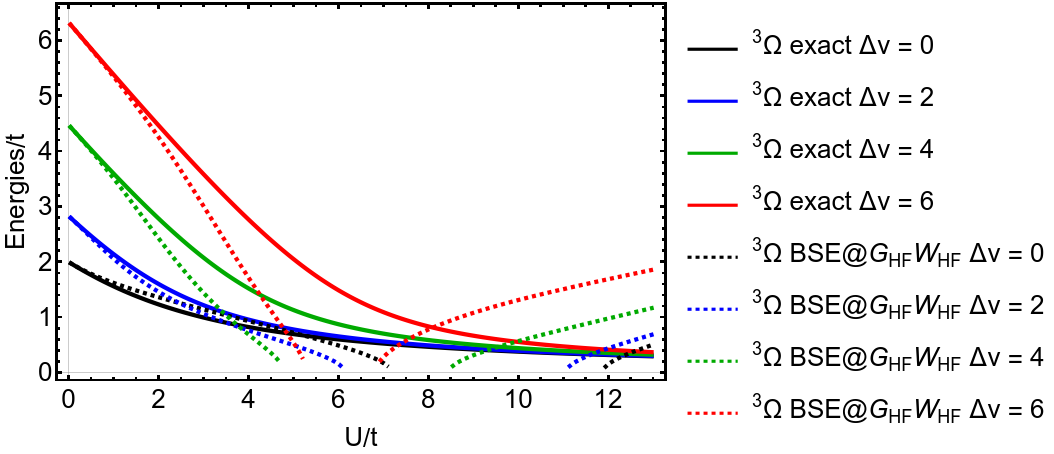}
    \includegraphics[scale=0.36]{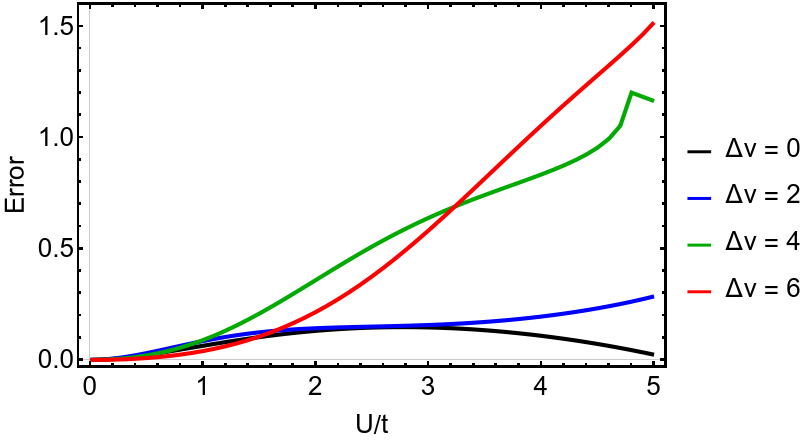}
    \includegraphics[scale=0.36]{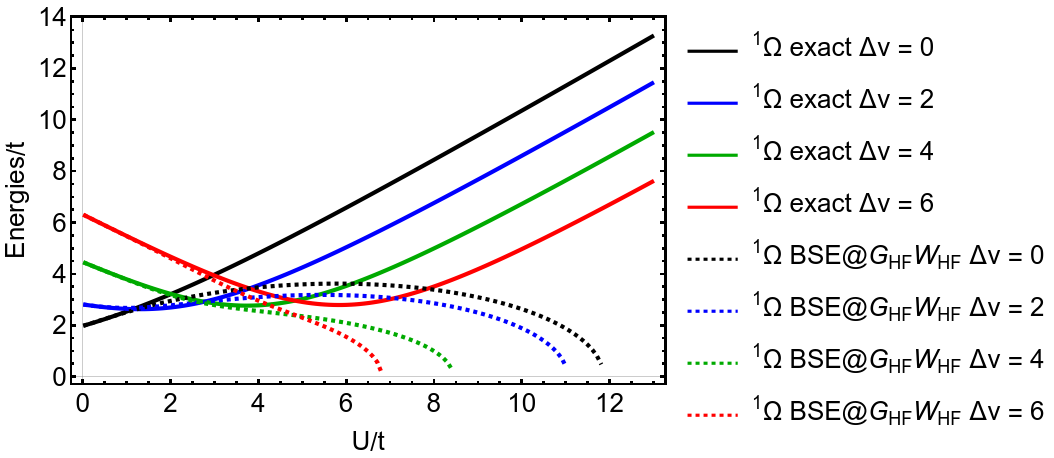}
    \includegraphics[scale=0.36]{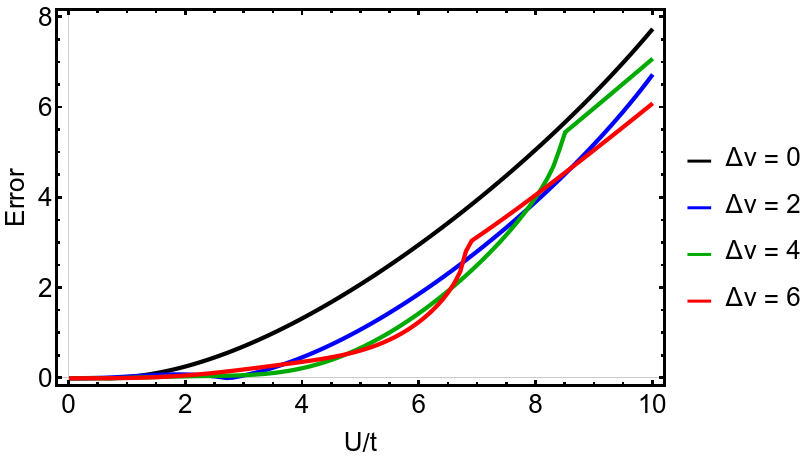}
    \caption{Triplet (top) and singlet (bottom) BSE@$G_\text{HF}W_\text{HF}$ neutral excitations (left) and their corresponding error with respect to the exact results (right) as functions of $U/t$ for various values of $\Delta v/t$.}
    \label{fig:w GW}
\end{figure*}

\begin{figure*}
    \includegraphics[width=0.49\linewidth]{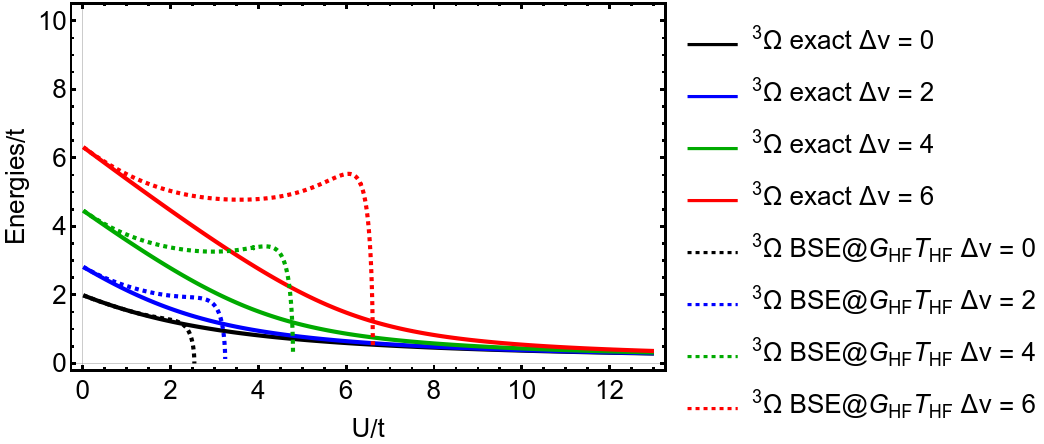}
    \includegraphics[width=0.49\linewidth]{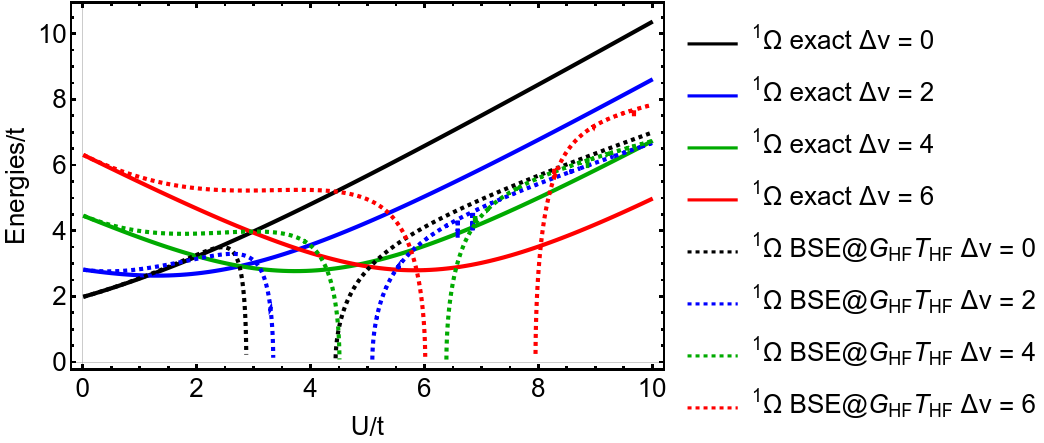}
    \caption{Triplet (left) and singlet (right) BSE@$G_\text{HF}T_\text{HF}$ neutral excitations as functions of $U/t$ for various values of $\Delta v/t$.}
    \label{fig:w GT}
\end{figure*}

We start by analyzing the triplet excitation. The triplet excited-state wave function $\ket*{{}^3\Psi^{N}}$ [see Eq.~\eqref{Psi dimer as}] does not depend on $\Delta v$, unlike the singlet ground-state wave function $\ket*{{}^1\Psi_0^{N}}$. 
At $\Delta v = 0$ and $U=0$, $\ket*{{}^1\Psi_0^{N}}$ is a linear combination of $\ket{\uparrow_{1} \downarrow_{2}}$, $\ket{\downarrow_{1} \uparrow_{2}}$, $\ket{\uparrow\downarrow_{1}0_{2}}$, and $\ket{0_{1}\uparrow\downarrow_{2}}$ with equal weights. 
The contribution from double occupancies, $\ket{\uparrow\downarrow_{1}0_{2}}$ and $\ket{0_{1}\uparrow\downarrow_{2}}$, decreases by increasing $U$ and increases by increasing $\Delta v/t$, as evidenced from Fig.~\ref{fig: Poids}. 
This excitation energy is well described by $GW$ up to $U/t \approx 5$ for $\Delta v=0$, as shown in Fig.~\ref{fig:w GW} (top left panel); the error with respect to the exact result increases with $\Delta v/t$ (see top right panel of Fig.~\ref{fig:w GW}). In the case of the singlet excitation energy, instead, we observe that the $GW$ error decreases by increasing $\Delta v/t$ (see bottom panels of Fig.~\ref{fig:w GW}).
For the $T$-matrix the results are shown in Fig.~\ref{fig:w GT}. For both the triplet and singlet transition energies, the error increases with $\Delta v$.

\par

\begin{figure*}
\includegraphics[width=\linewidth]{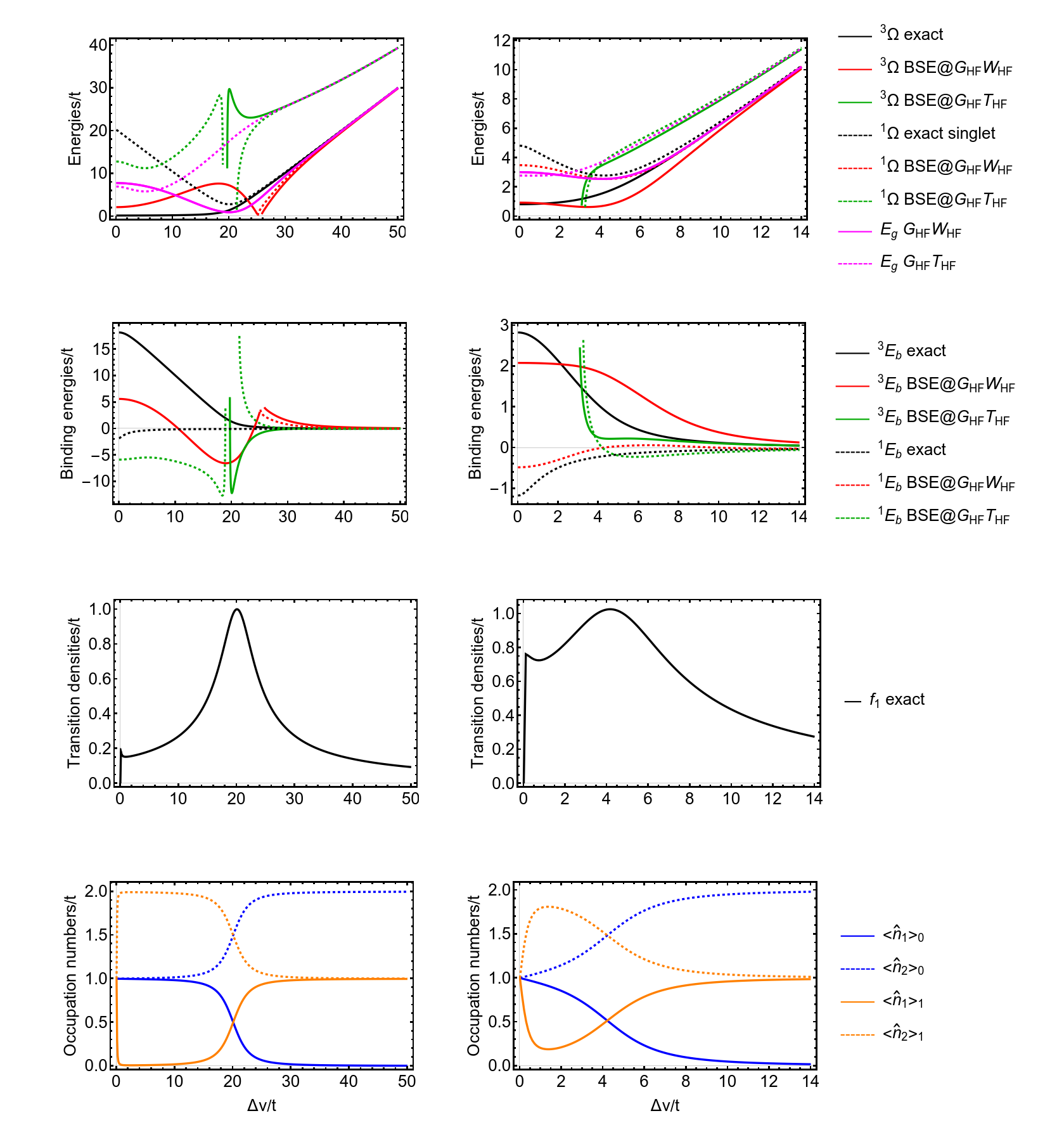}
    \caption{Neutral excitations (first row), binding energies (second row), transitions densities (third row), and occupation numbers (fourth row) of the ground state and the first singlet excited state as functions of $\Delta v/t$. 
    Results obtained for $U/t = 20$ (left column) and $U/t = 4$ (right column).}
    \label{fig:grille_GW}
\end{figure*}

In Fig.~\ref{fig:grille_GW}, we analyze in more detail the trend of the singlet and triplet excitation energies as a function of $\Delta v/t$. 
First, we notice that, since the singlet ground- and excited-state wave functions depend on $\Delta v/t$, the transition probabilities to the singlet states also depend on $\Delta v/t$. 
In particular, the exact transition probabilities
\begin{equation}
    f_{1}=\mel*{{}^1\Psi^{N}_{0}}{\hat{n}_{1}-\hat{n}_{2}}{{}^1\Psi^{N}_{1}}
\end{equation}
associated with the singlet excitation is peaked at $\Delta v/t=U/t$, as shown in the second-last panel of Fig.~\ref{fig:grille_GW}, which goes with a change in the site occupation numbers of the ground state and the corresponding singlet state ($\expval*{\hat{n}_{i}}_{0} \equiv \mel*{{}^1\Psi^{N}_{0}}{\hat{n}_{i}}{{}^1\Psi^{N}_{0}} $ and $\expval{\hat{n}_{i}}_{1} \equiv \mel*{{}^1\Psi^{N}_{1}}{\hat{n}_{i}}{{}^1\Psi^{N}_{1}}$, respectively, with $i = 1$  or $2$, indicating the site). We observe that, precisely at the value $\Delta v/t=U/t$, the singlet excitation energy calculated within the $GT$ approximation encounters a singularity in the case $U/t=20$, whereas, for $U/t=4$, the singularity disappears but the energy becomes complex below $\Delta v/t=4$. 
The triplet excitation energy, instead, becomes complex below $\Delta v/t=U/t$. 
In the case of the $GW$ approximation, the singlet excitation energy is complex over a slightly larger range of $\Delta v/t$ for $U/t=20$, whereas it remains always real for $U/t=4$ (see first panel of Fig.~\ref{fig:grille_GW}). 
Overall, for $\Delta v/t \ll U/t$, $GW$ describes better than $GT$ the triplet excitation energy for $U/t=20$ and both the singlet and triplet transitions for $U/t=4$. 
The corresponding binding energies, however, suffer from the $GW$ error in the fundamental gap.
Of course, the problem observed in the approximate BSE in the stongly correlated regime is partially inherited from the incorrect description of the quasiparticle energies in this limit. 
At the level of the BSE kernel, one could explore the effect of second-order terms in $W$ or in $T$, which are usually neglected, or go beyond the static approximation, which may not be justified in case of low-energy excitations. \cite{Strinati_1988}

For $\Delta v/t \gg U/t$, the exact and approximate kernels of the BSE become negligible so that the triplet and singlet excitation energies tend to the fundamental gap: HF and $GW$ excitation energies merge to the exact ones since the HF and $GW$ fundamental gaps tend also to the exact gap for large $\Delta v/t$ (see Fig.~\ref{fig:Gap Delta v}); instead, the $GT$ excitation energies tend to the $GT$ fundamental gap, which, as we discussed above, deviates from the exact one. Of course, these trends in the excitation energies influence the quality of the corresponding binding energies, which tend to the exact results in this limit.


\par

\section{Conclusions}
\label{sec:conclusions}

Based on the exactly-solvable asymmetric Hubbard dimer model, we have gauged the accuracy of the charged and neutral excitation energies in different correlation regimes obtained with two distinct approximations of the self-energy: the $GW$ and $T$-matrix approximations which correspond to a resummation of different families of diagrams.
In particular, using these two approximate self-energies and their corresponding kernel, excited-state energies were computed within the Bethe-Salpeter formalism.

Overall, we have found that the $GW$ approximation works better than the $GT$ approximation both for the quasiparticle energies and the neutral excitation energies as functions of the degree of correlation ($U/t$) and asymmetry ($\Delta v/t$) in the system. 
In particular, the $GT$ quasiparticle energies do not exhibit the correct behavior as $\Delta v/t \to \infty$. 
In this limit, correlation becomes negligible in the exact case and in $GW$, but not in $GT$, because of the non-vanishing particle-particle polarizability which enters into the $T$-matrix expression. 
Because of this, the behavior of the neutral excitation energies is also incorrect in the large $\Delta v/t$ limit, unlike their HF and $GW$ counterparts. 
This suggests that the quality of the $GT$ results can be sensitive to inhomogeneities in real materials. 
To shed more light on this issue it would be interesting to analyze this issue in real materials. \cite{Muller_PRB2019,Nabok_2021}
Moreover, $GT$ seems to be more sensitive to the starting Green's function, at least for the Hubbard dimer, and this can strongly affect the quality of the results.
Finally, the strongly correlated regime remains a challenge for both approximations studied in the present manuscript.

\acknowledgments{
This project has received financial support from the Centre National de la Recherche Scientifique (CNRS) through the 80$\vert$PRIME program and has been supported through the EUR grant NanoX under Grant No.~ANR-17-EURE-0009 in the framework of the ``Programme des Investissements d’Avenir''.
PFL also thanks the European Research Council (ERC) under the European Union's Horizon 2020 research and innovation programme (Grant agreement No.~863481) for funding.}

\bibliography{references}

\end{document}